\documentclass[sn-mathphys]{sn-jnl}
\jyear{2021}
\usepackage{xcolor}
\usepackage{amsmath}
\usepackage{subfigure}
\usepackage{xparse}
\theoremstyle{thmstyleone}
\theoremstyle{thmstyletwo}
\theoremstyle{thmstylethree}

\makeatletter
\NewDocumentCommand{\LeftComment}{s m}{\Statex \IfBooleanF{#1}{\hspace*{\ALG@thistlm}}\(\triangleright\) #2}
\makeatother
\begin{document}
\title[Ripple Knowledge GCN For Recommendation Systems]{Ripple Knowledge Graph Convolutional Networks For Recommendation Systems}

\author[1]{\fnm{Chen} \sur{Li}}
\author[2]{\fnm{Yang} \sur{Cao}}
\author[2]{\fnm{Ye} \sur{Zhu}}
\author[3]{\fnm{Debo} \sur{Cheng}}
\author[4]{\fnm{Chengyuan} \sur{Li}}
\author[4]{\fnm{Yasuhiko} \sur{Morimoto}}

\affil[1]{\orgdiv{Graduate School of Informatics}, \orgname{Nagoya University}, \orgaddress{\city{Chikusa, Nagoya} \postcode{464-8602},  \country{Japan}}}
\affil[2]{\orgdiv{Centre for Cyber Resilience and Trust}, \orgname{Deakin University}, \orgaddress{\city{Burwood} \postcode{3125},  \country{Australia}}}
\affil[3]{\orgdiv{Science, Technology, Engineering and Mathematics (STEM)}, \orgname{University of South Australia}, \country{Australia}}
\affil[4]{\orgdiv{Graduate School of Engineering}, \orgname{Hiroshima University}, \orgaddress{\city{Higashi-hiroshima} \postcode{10587},  \country{Japan}}}
 
\abstract{Using knowledge graphs to assist deep learning models in making recommendation decisions has recently been proven to effectively improve the model’s interpretability and accuracy. This paper introduces an end-to-end deep learning model, named RKGCN, which dynamically analyses each user’s preferences and makes a recommendation of suitable items. It combines knowledge graphs on both the item side and user side to enrich their representations to maximize the utilization of the abundant information in knowledge graphs. RKGCN is able to offer more personalized and relevant recommendations in three different scenarios. The experimental results show the superior effectiveness of our model over 5 baseline models on three real-world datasets including  movies, books, and music.}

\keywords{Deep Learning, Recommendation Systems, Knowledge Graph, Graph Convolutional Networks, Graph Neural Networks
\footnotetext{Link:~\href{https://link.springer.com/article/10.1007/s11633-023-1440-x}{}}
\footnotetext{Machine Intelligence Research, Vol.21(3).  DOI:~\href{10.1007/s11633-023-1440-x}{}
}}
\maketitle

\section{Introduction}
\label{sec:introduction}
With the current development of big data and internet technology, most daily life products are linked through {social networks and mediums} to facilitate business and advertising \cite{zhang2022big}. However, the massive data growth also leads to information overload problems. The vast amount of information available on the internet can be overwhelming for users, making it difficult for them to find and select items that align with their interests and needs among the numerous options available. 
To alleviate the information overload problem, recommendation systems have been designed to assist users in finding and selecting items of interest~\cite{afsar2022reinforcement,wu2022graph}.

\begin{figure}[htpb]
\centering
\includegraphics[width=\textwidth, height=4.7cm]{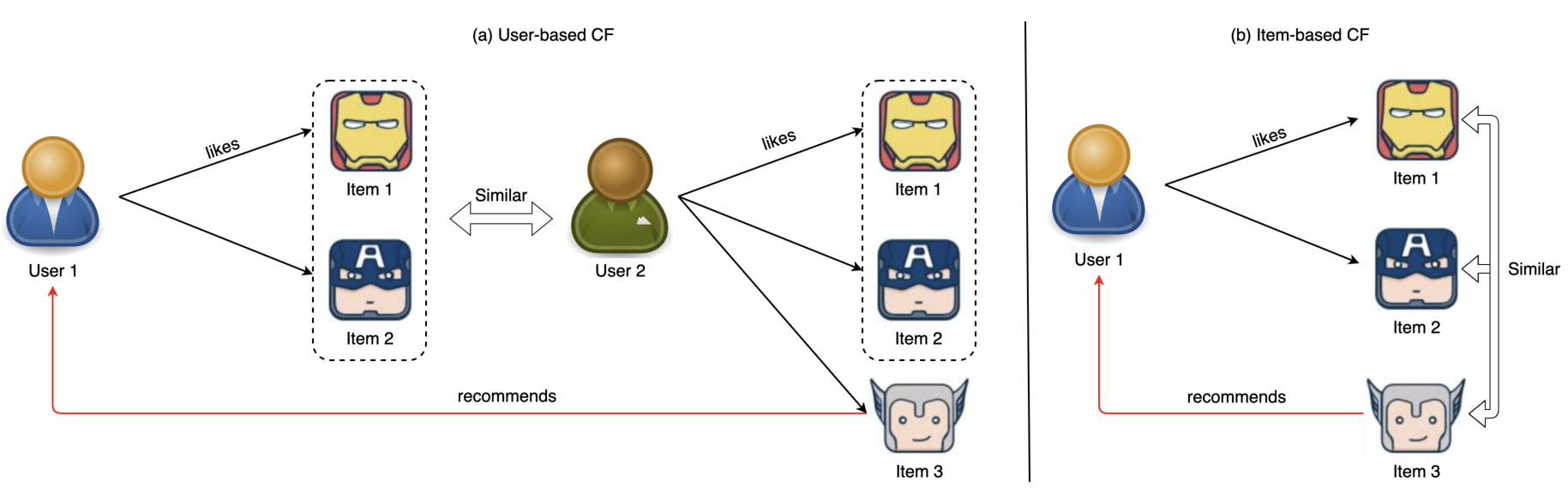}
\caption{{An example of a recommendation system for CF.}}
\label{fig:cf}
\end{figure}

\begin{figure}[t]
\centering
\begin{subfigure}[An example of a rating matrix.]{
\includegraphics[width=0.7\hsize]{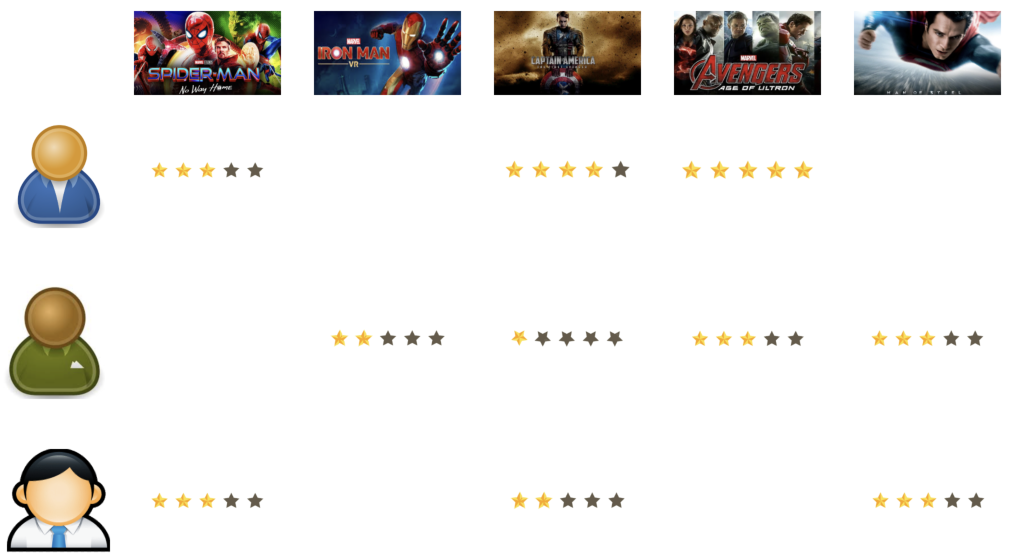}
}
\end{subfigure}
\begin{subfigure}[An example of a purchase/click matrix.]{
\includegraphics[width=0.7\hsize]{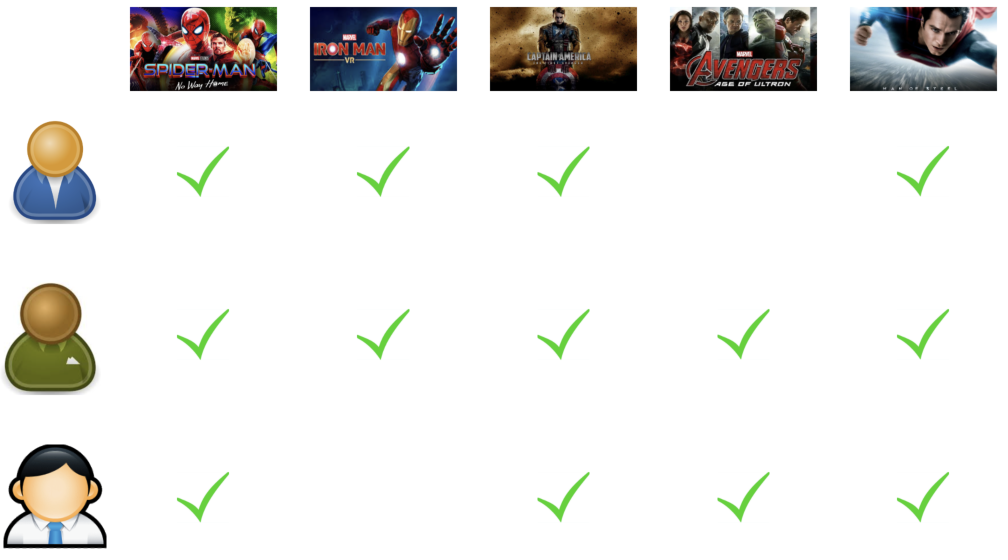}
}
\end{subfigure}
\caption{{Examples of rating and purchase/click matrices for a movie recommendation system.}}
\label{fig:rating}
\end{figure}

{Collaborative filtering} (CF) is a traditional approach of a recommendation system that represents users and items as {vectors} and models {user-item} interactions through {concrete actions}, such as an inner product {operation} \cite{wang2017joint} or a neural network \cite{he2017neural}. {Generally, CF basically has two approaches, i.e., user- and item-based CFs. Figure \ref{fig:cf} demonstrates an example of a recommendation system for CF. User-based CF (Fig. \ref{fig:cf} (a)) is based on the similarity of users or neighborhoods. Item-based CF (Fig. \ref{fig:cf} (b)), on the other hand, is based on the similarity between items. Usually, CF can be considered the task of completing a sparse user-item rating matrix. In other words, the main task is the prediction of an absence of ratings. Figure \ref{fig:rating} (a) illustrates a rating matrix based on the CF approach. The rating matrix is explicit feedback from users on their movie preferences. However, explicit feedback is sparse or not always available in many applications \cite{ma2022vae++}. For example, in the case of e-commerce websites, many users do not rate the items they have purchased. Therefore, recommendation systems tend to use implicit feedback such as purchase history, clickstream, and browsing activity to generate recommendations. Figure \ref{fig:rating} (b) shows a case of a purchase matrix, where a checkmark represents a user action such as a purchase or a click.} 

However, {the user-item interactions for the CF-based approaches are sparse thus having the limitation of a “cold start" problem}, where the system has little or no information about a new user or item~\cite{wang2019knowledge}. {To overcome these limitations, most researchers use feature-rich scenarios in which user and item attributes are used to compete for sparsity and improve recommendation system performance.} They also recognized that attributes are not isolated, but rather are interconnected and form a knowledge graph. {Essentially, a knowledge graph is a semantics-related network, which is a graph structure that consists of nodes and edges } \cite{wang2014knowledge, nickel2015review}. {Each node and edge in the knowledge graph represent a real-world entity and a relationship between entities, respectively}~\cite{ehrlinger2016towards}. {A knowledge graph can be employed to provide extra contextual information} and structure for the recommendation model, allowing it to better understand the characteristics and relationships of the users and items \cite{wu2022survey}. This can help the model make more informed recommendations and improve the overall accuracy of the system.

\begin{figure}[htpb]
\centering
\includegraphics[width=\textwidth]{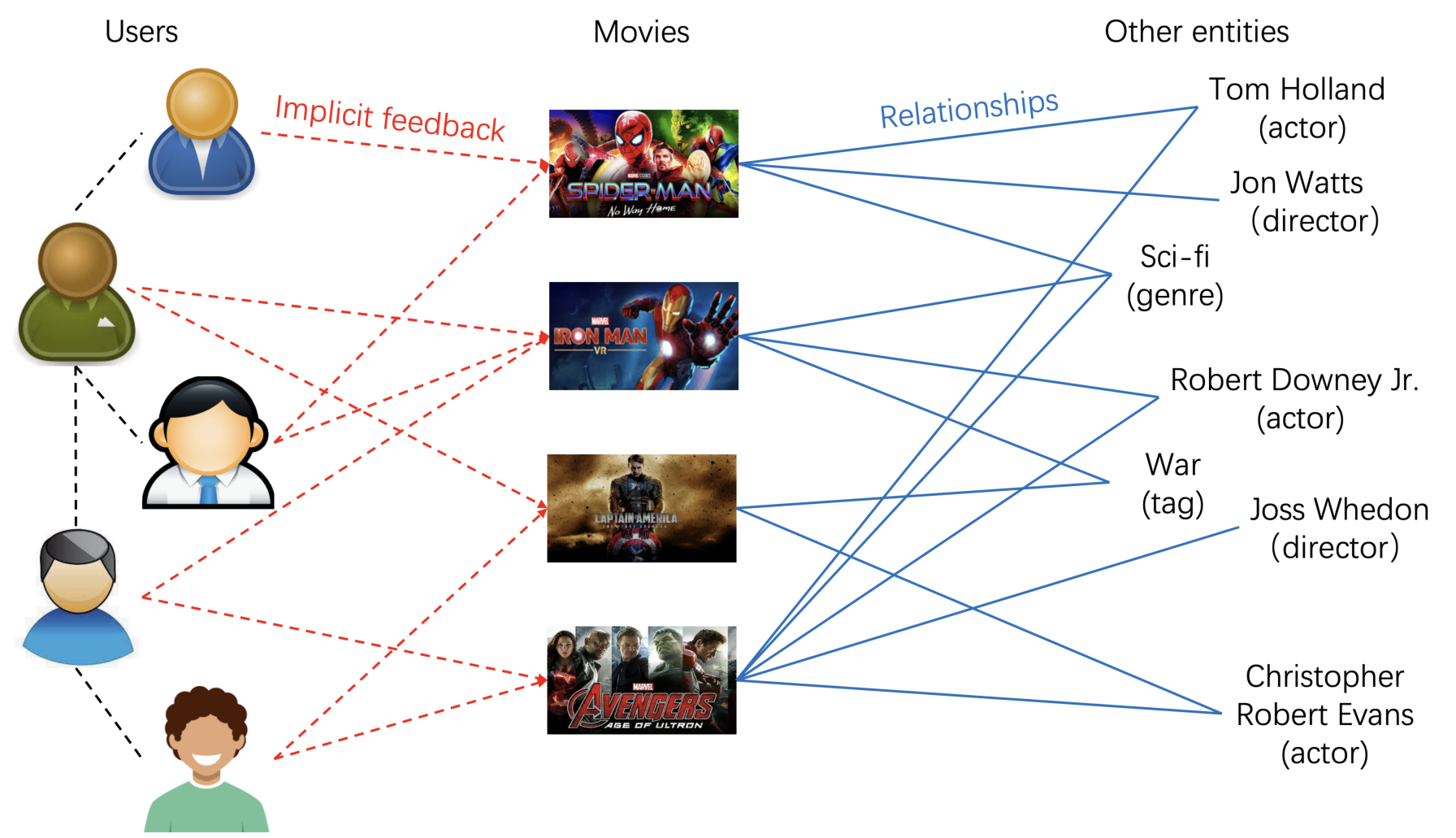}
\caption{{An example of a knowledge graph-enhanced movie recommendation system.}}
\label{fig:kg}
\end{figure}

Deep learning can be used to build and refine knowledge graphs by using {extensive unstructured and structured data} to learn about the relationships between different entities and their properties \cite{chen2020review}. {Figure \ref{fig:kg} shows an example of an enhanced movie recommendation system with knowledge graphs. Knowledge graphs provide a rich connection between users, movies, actors and directors. A user's preferences may be affected by similar genres, other movies in which the same actors have acted, or other movies directed by the same director. Therefore, rich knowledge graph information helps improve the accuracy, diversity and interpretability of the suggested recommendation results.} There are several types of deep learning models that are commonly used for knowledge graph representation and reasoning, which can {be organized into three main classes}:

\begin{itemize}

\item knowledge graph embedding models~\cite{wang2017knowledge}: represent the nodes and relationships in a knowledge graph as low-dimensional vectors (embedding) in a continuous vector space. The embeddings are learned such that they preserve the structure and properties of the knowledge graph, and can be used for various downstream tasks such as link prediction, triplet classification, and entity resolution. Examples of knowledge graph embedding models include~\cite{bordes2013translating}, TransR~\cite{bordes2013translating}, and ComplEx~\cite{trouillon2016complex}. 
    
\item Graph neural networks (GNNs)~\cite{kipf2016semi}: operate on graph-structured data and use neural networks to learn the representations of the nodes and edges in the graph. GNNs can be used for tasks such as node classification, graph classification and graph generation. There are many different types of GNNs, including \emph{graph convolutional networks} (GCNs), \emph{graph attention networks} (GATs), and graph autoencoders.
    
\item Rule-based approaches~\cite{chen2020review}: use logical rules to perform reasoning over a knowledge graph. The rules are typically encoded as Horn clauses, and the model uses a rule-based engine to perform logical inferences over the knowledge graph. Examples of rule-based approaches include \emph{probabilistic soft logic} (PSL) and \emph{Markov logic networks} (MLNs).

\end{itemize}

Existing knowledge graph recommendation systems, such as RippleNet~\cite{wang2018ripplenet} and KGCN~\cite{wang2019knowledge} incorporate the user's interaction history into the knowledge graph, allowing the system to simulate the user's preferences based on their past interactions. However, RippleNet does not take into account the relationships between different items in the interaction history, and thus may not be able to fully capture the user's dynamic preferences when facing different candidate items. In addition, KGCN only focuses on the representation of items and ignores the fact that the knowledge graph can also be used to enrich the representation of users and their preferences. This may limit the system's ability to provide personalized and relevant recommendations.

To address the limitations of existing recommendation models, this paper proposes an end-to-end deep learning model called \emph{representation-enhanced knowledge graph convolutional networks}(RKGCN) to predict the click-through rate (CTR) of users. The core idea of RKGCN is to use the knowledge graph to simultaneously improve the representation of both users and items, and then utilize the enhanced representations to capture the dynamic interests of users when they encounter different items, in order to make more accurate recommendations. By using the knowledge graph to enrich the representation of both users and items, RKGCN is able to offer more personalized and relevant recommendations. {The main contributions of this study} are summarized as follows:

\begin{itemize}

\item {{\bf An efficient knowledge graph-based neural network:}} Proposing an end-to-end framework based on knowledge graph convolutional networks for recommendation systems in graph data.
    
\item {{\bf Refinement of the effect of different items on user preferences:}} Combining two different methods, {i.e., the RippleNet and KGCN}, to improve the performance of the recommendation system. 

\item {{\bf Validation on multiple benchmark datasets:}} Evaluating the performance of RKGCN on three real-world datasets. Our empirical results show that RKGCN performs better than 5 state-of-the-art methods.

\end{itemize}

The remainder of the paper is organized as follows. We first describe the related work of knowledge graph-based neural networks in Section~\ref{sec:related}. {Next, the problem for change-interval detection is defined} in Section~\ref{sec:prob}. {Then, our proposed RKGCN is described in detail in Section \ref{sec:rkgcn}.} The experimental settings and results are presented in Section~\ref{sec:exp}. Finally, the concluding remarks are given in Section {\ref{sec:conclusion}}.

{
\section{Related Work}
\label{sec:related}
In this section, we first review conventional recommendation system algorithms such as CF. However, due to the fundamental limitations of CF methods in constructing user-item matrices, we then discuss neural network-based techniques that have become popular in recent years. Finally, some advanced graph-based neural networks for recommendation systems are surveyed.

\subsection{Collaborative Filtering for recommendation Systems}
CF is the most commonly used approach for recommendation systems. The fundamental assumption of CF is that if two users behave similarly on items (e.g., purchase, watch, listen), then similar ratings or actions are performed on other items by default \cite{goldberg2001eigentaste}. Two users are similar if they rate a similar set of items, which is called a user-based approach \cite{bell2007improved}. Similarly, an item-based approach calculates user preferences for an item based on users' ratings of similar items \cite{sarwar2001item}. Generally, we often use latent factor models to map users and items into vectors of latent factor space and then make similar recommendations or predictions. The user and item are directly comparable in this space. Most latent factor-based models factorize the rating matrix using single value decomposition (SVD) \cite{paterek2007improving}. However, CF approaches usually inevitably have issues such as interactive sparsity and cold start, which affect the accuracy of recommendations.

\subsection{Neural Networks in Recommendation Systems}
Neural networks have been successful in image classification, speech recognition, machine translation, and many other research fields \cite{smys2020survey}. Inspired by their powerful capabilities, many research works have applied neural networks to recommendation systems. Restricted Boltzmann machines (RBMs) were used to handle recommendation tasks \cite{salakhutdinov2007restricted}. As an extension of RBMs, a collaborative filtering model based on a stacked denoising autoencoder and sparse inputs with latent factors is proposed \cite{strub2015collaborative}. The model has two steps, encoding and decoding. The stacked denoising autoencoder extracts the latent factors of users and items. Then, the rating matrix is completed by a decoder. RNNs are a useful tool for sequence prediction and classification, which have been used in recommendation systems. For example, RNNs have been successfully applied to quote recommendations for dialogue and writing \cite{ahn2016quote}. That work focused on recommending proverbs and quotations to users who are writing based on the body of the current text. Additionally, recommendation systems that predict user purchase behavior from the content of tweets have been developed \cite{korpusik2016recurrent}. Developers believe that previous tweets have an impact on subsequent tweets and that tweets are effective indicators of users' future purchase behavior. They used the RNN model to model the continuous tweets of users and predict their purchase preferences. 

Moreover, session-based recommendations have been proposed for sequential data \cite{li2018capturing}. Sessions are sequences of pages sorted by timestamps, such as pages clicked by users. A session-based recommendation system aims to generate real-time recommendations based on session data rather than historical rating data. The session-based approach mainly recommends to users the items they are currently searching for (short-term preferences). Since the length of a session is variable, DRNNs with multiple hidden layers can be trained in a session-parallel manner \cite{hidasi2015session}. In the training phase, the loss is measured in terms of ranking loss, such as TOP1 \cite{hidasi2015session} and Bayesian personalized ranking \cite{rendle2012bpr}.

\subsection{Graph Neural Networks in Recommendation Systems}
In recent years, heterogeneous graph-based recommendation systems have made significant achievements. The representation of entities in heterogeneous graphs can be effectively improved by analysing the relationships between nodes in heterogeneous graphs. Moreover, recommendation systems that utilize structural information \cite{feng2012incorporating} and the contextual information of entities \cite{zhang2016collaborative} have been proposed. Furthermore, personalized recommendations are made using additional social information \cite{wang2016social} and location information \cite{yin2013lcars} of users. In addition, different processing of knowledge graphs can classify graph-based recommendation systems into path-based \cite{hu2018leveraging} and node-based \cite{wang2019knowledge} approaches. 

KGCN \cite{wang2019knowledge} leveraged a graph-based convolutional neural network to model information transfer from edge nodes to central nodes in a knowledge graph. KGCN is a special graph convolutional network. It treated the nodes closest to the central node as the primary source of information while assigning different weights to different relationships on the graph convolutional neural network during updating. RippleNet is a hybrid model based on both the path- and node-based approaches \cite{wang2018ripplenet}. The model propagates personalized recommendations to users by propagating users' preferences over the knowledge-edge graph. RippleNet assumes that users' preferences are spreading outwards. The corresponding relationships in the graph are from other entities of the history. Therefore, historical entities closer to the user are more representative of current preferences in a knowledge graph. RippleNet obtained users' latent representations by computing the similarity between the candidates of corresponding relationships in the space and those entities that are nearer to the user. Finally, the information of the embedded user representations becomes rich with the propagation on the knowledge graph.

RippleNet and KGCN incorporate the interaction history of users into a knowledge graph, enabling the system to model the preferences of users according to their previous interactions. However, RippleNet does not consider the relationship between different items in the interaction history and cannot fully capture the dynamic preferences of users. KGCN only considered item representations without a knowledge graph, which could enrich users' preferences. Therefore, it has a limited ability to provide personalized and relevant recommendations. Inspired by RippleNet and KGCN models, this paper proposes an RKGCN that uses knowledge graphs to improve both user and item representations and then uses the enhanced representations to capture the dynamic interests of users. By using knowledge graphs to enrich the representations of users and items, RKGCN can provide more personalized and relevant recommendations.
}

\section{Knowledge Graph Convolutional Networks}
\label{sec:prob}

 {Our proposed model is inspired by the non-spectral graph convolutional network (NSGCN) \cite{niepert2016learning}. Generally, the NSGCN approach acts on original graphs and defines convolution for special types of graphs (i.e., knowledge graphs).} This section provides a basic definition of the knowledge graph, user preference set, entity neighbor set, and the formula for the CTR prediction task.  {For a better understanding of the proposed model, table \ref{tab:summary} summarizes the notations used in this paper.}

\begin{table}[ht]
\renewcommand{\arraystretch}{1.5}
\caption{ {Summary of notations used in this paper.}}
\centering
{ 
\begin{tabular}{ll}\hline
Notation & Meaning\\\hline
$U$ & the set of $m$ users\\
$u_i$ & the $i$-th user in the user set $U$\\
$V$ & the set of $n$ items\\
$v_j$ & the $j$-th item in the item set $V$\\
$Y$ & the user-item interaction matrix\\
$y_{u,v}$ & the element of $u$-th user and $v$-th item of the user-item interaction matrix $Y$ \\
$\mathcal{F}$ & the prediction function of a user's interaction with an item\\
$\Theta$ & model parameters\\
$G$ & knowledge graphs\\
$h$ & the head of a knowledge graph\\
$r$ & the relation of a knowledge graph\\
$t$ & the tail of a knowledge graph\\
$N(v)$ & the neighbor set of item $v$\\
$P_u^k$ & the preference set of user $u$ with item $k$\\
$P_u^0$ & the histroical interests set of user $u$\\
$S_u^k$ & the Similarity set of user $u$\\
$R$ & the realtion embedding in $S_u^k$\\
$a_i$ & the similarity of the candidate items\\
$O_u^1$ & the 1-stage representation of the user \\
$o_n$ & the user's embedding \\
$\text{score}_u^r$ & the correlation score between user $u$ and relation $r$ \\
$N_u^v$ & the neighbor information of the current item $v$ \\
$\hat{y}_{u,v}$ & the prediction of user $u$ \\
$T_u$ & the number of negative samples \\
$\mathcal{C}$ & the cross entroy function \\
$\lVert\mathcal{C}\rVert_2^2$ & the L2-regularizer of $\mathcal{C}$\\\hline
\end{tabular}
}
\label{tab:summary}
\end{table}

\subsection{Problem Formulation}
Let $U=\{u_1, u_2, ..., u_m \}$ and $V=\{v_1, v_2, ..., v_n \}$ indicate a set of $m$ users and $n$ items, respectively. Thus, the user-item interaction matrix can be described as $Y=\{y_{u,v} \| u \in U, v \in V\}$. Usually, the matrix $Y$ is created from the implicit interaction information. If a user interacts with an item (e.g., clicks, watches, listens), the value in the matrix is 1. Otherwise, it is 0. The equation is shown as follows:

\begin{equation}
\label{eq:1}
y_{u,v}= \left\{
\begin{array}{lr}
1,~~\text{if observed interation between $u$ and $v$}&\\
0, ~~\text{otherwise}& 
\end{array}
\right.
\end{equation}

Knowledge graphs can be enriched with information on users' preferences in recommendation systems. A recommendation system aims to learn a function $\mathcal{F}$ to predict the probability of a user's interaction with an item, given known model parameters $\Theta$, true labels $Y$, and knowledge graphs $G$. The equation is shown below:

\begin{align}
\label{eq:2}
\hat{y}_{u,v}=\mathcal{F}(u, v \mid \Theta, Y, G).
\end{align}

A knowledge graph is a combination of many triples $(h, r, t)$, where $h$ and  {$r$ denote} the head and relation of a knowledge triple, respectively, while $t$ is the tail.  For example, the triplet \textit{(Titanic, Film.actor, Leonardo Wilhelm DiCaprio)} indicates that the movie \textit{Titanic} is acted by \textit{Leonardo Wilhelm DiCaprio}. \textit{Titanic} is the head entity $h$, \textit{Film.actor} is the relation $r$, and \textit{Leonardo Wilhelm DiCaprio} is the tail entity $t$.

\subsection{User preference set}
\begin{figure}[htpb]
\centering
\includegraphics[width=0.8\textwidth]{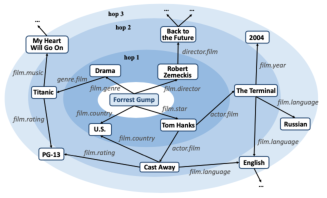}
\caption{{An example of user’s preference set.}}
\label{fig:preference}
\end{figure}

In the model RippleNet \cite{wang2018ripplenet}, the user’s preferences are simulated by the multi-stage propagation of user history, and we used this setting, Figure {\ref{fig:preference} shows a simple example of a 3-stage preference set of a user whose click history only has the Forrest Gump movie}. We assume that a user has only watched one movie, Forrest Gump, then the user’s 1-stage preference set $P^1$ includes four entities: Drama, U.S, Tom Hanks, and Robert Zemeckis. Similarly, the entity contained in the hop2 part of the figure is the user’s 2-stage preference set $P^2$. {In reality, the number of items in the interaction history of a user is variable.} Since various users have different sizes of their preference sets, we follow the strategy proposed in \cite{wang2019knowledge}, i.e., using a fixed-size sampler and the size of the sampler is an adjustable parameter.

\subsection{Entity Neighbour Set}
The definition of the entity neighbor set is similar to the definition of the user preference set. However, for the entity neighbor set, there is no need to consider the user’s history; {only the entities in the knowledge graph and the item $v$ that is directly connected belong to the neighbor set $N(v)$}.

 {The neighbor entities of this candidate item also have their own neighbor sets. Hence, aggregation can be used to realize the diffusion of neighbor information from outside to inside. Figure \ref{fig:hop} is an example of aggregation from the second hop to the candidate item entity. The elements of the neighbor set of Iron Man are Sci-fi and the actor Robert Downey Jr. 

\begin{figure}[htpb]
\centering
\includegraphics[width=0.45\hsize]{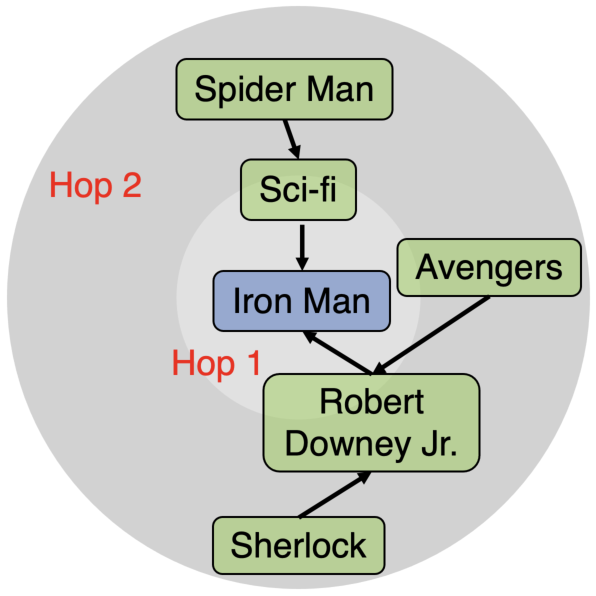}
\caption{{An example of an entity update. An entity embedding is updated from peripheral nodes to central nodes by a loop operation.}}
\label{fig:hop}
\end{figure}
}

\section{Proposed method}
\label{sec:rkgcn}
The content of this chapter introduces the two components of the RKGCN model and their formulas in detail and customizes the loss function for optimization.

\subsection{Overview of RKGCN}
\begin{figure}[htpb]
\centering
\includegraphics[width=\textwidth, height=7cm]{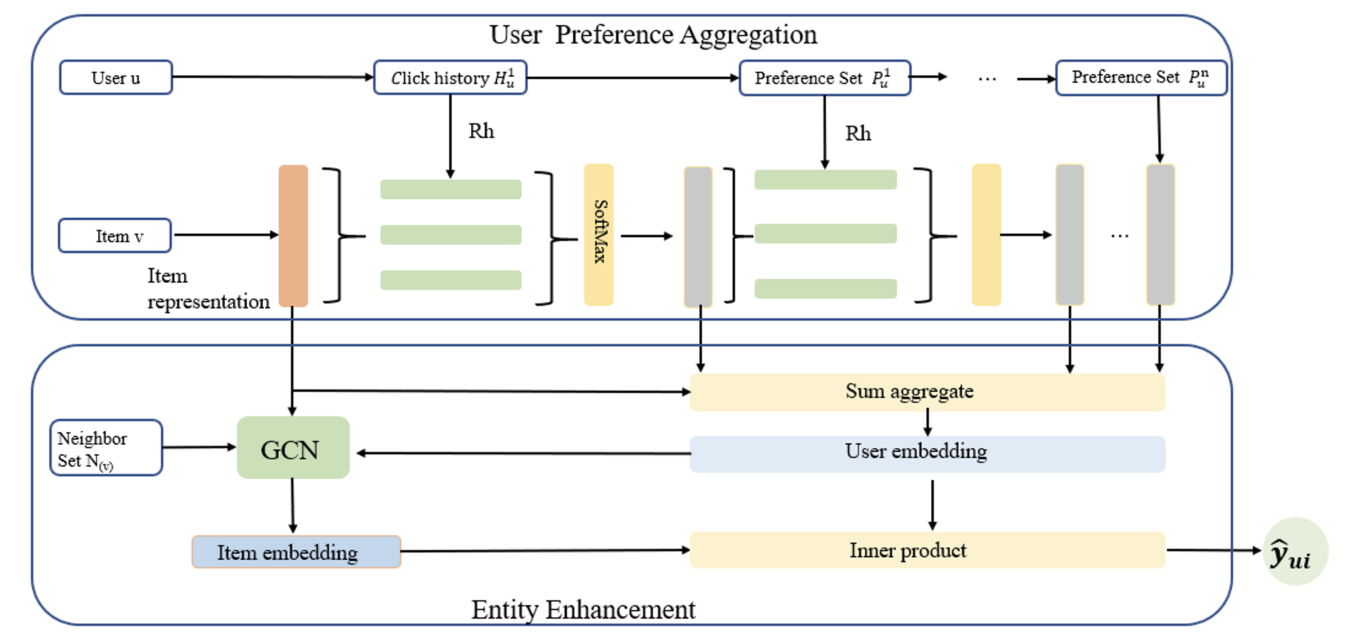}
\caption{{Overview architecture of the proposed RKGCN.}}
\label{fig:rkgcn_overview}
\end{figure}

Figure \ref{fig:rkgcn_overview} shows an overview of the RKGCN. We can see that the RKGCN has two main components: user preference aggregation and physical neighbor set. Finally, we obtain the user and item embeddings and then use the inner product to {calculate the probability. The details of the model are introduced in the next section}.

\subsection{User Preference Aggregation}
Similar to the traditional collaborative filtering {model \cite{su2009survey}}, we believe  that users will be more inclined to choose similar items in their interaction history to interact. In the knowledge graph, the similarity between items is reflected in their connectivity. The closer the distance between entities, the more similar they are. The user preference aggregate is based on this principle, representing the user as the aggregation of {entities closest to the interaction history of the user in a knowledge graph}. First, we need to obtain the user’s preference set $P_u^k, (k = 1, 2...H)$. {The equation is shown below:

\begin{align}
\label{eq:3}
P_u^k=\{t \mid (h,r,t \in K~~\text{and}~~h \in P_u^{k-1})\},~~(k=1,2,\cdots,H),
\end{align}
where $P_u^0$ is the historical interests set of user $u$. 

To obtain the similarity between the items in the user’s historical click record and the candidate item, we also calculate another set $S_u^k$. $S_u^k$ is shown as follows:

\begin{align}
\label{eq:4}
S_u^{k}=\{(h,r,t) \mid (h,r,t \in K~~\text{and}~~h \in P_u^{k-1})\},~~(k=1,2,\cdots,H),
\end{align}
where $S_u^k$ is slightly different from $P_u^0$. The elements in $S_u^k$ are the triples whose head entities are the elements of $P_u^k$. Then, the similarity of the candidate items to the items in the history of a user in the relation $R$ is $a_i$.

\begin{align}
\label{eq:5}
a_i=\text{Softmax}(Rhv^t)
\end{align}
where $R \in \mathbb{R}^{d\times d}$ denotes the relation embedding in the $S_u^k$. $h \in \mathbb{R}^d$ denotes the embedding of the head entity in $S_u^k$.  {$v^t$ is the transpose of an item embedding vector $v$.} The 1-stage representation of the user is represented by $O_u^1$, and 

\begin{align}
\label{eq:6}
O_u^1=\sum_{(h,r,t \in S_u^1)}a_i t_i.
\end{align}
We also aggregate the user’s n-stage representation. A sum aggregator can achieve the above functions.

\begin{align}
\label{eq:7}
o_n = W(o_{n-1}+v), ~~(n=1,2,\cdots, H),
\end{align}
where $v \in \mathbb{R}^d$ indicates that $v$ is a candidate item’s representation. The reason for adding candidate item representation to the aggregator is based on an assumption. When the same user faces different candidate items, the user’s embedding should be different and have a strong correlation with the candidate products. $o_n$ denotes the user’s embedding.
}

\subsection{Entity Enhancement}
\begin{figure}[htpb]
\centering
\includegraphics[width=0.9\hsize]{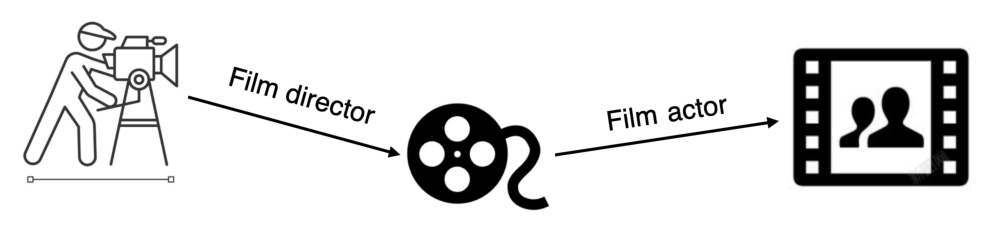}
\caption{{An example of the importance of relation score.}}
\label{fig:relation}
\end{figure}

After we obtain the user embedding enhanced by the knowledge graph, we can use KGCN to obtain the user’s preference for different relationships and then use the neighbor nodes in the knowledge graph to update the candidate item’s embedding. Figure \ref{fig:relation} is a simple example to illustrate the importance of relationship scores. Suppose the candidate item is the movie Iron Man. When we use the knowledge graph to enhance this entity’s embedding, the information of its neighboring entities has different weights for users. Suppose the current user prefers the starring role of this movie. In that case, the weight of the relation film.film.actor should be higher, and more information about the entity Robert Downey Jr will be added into the embedding of entity Iron Man 3. The inspiration for entity enhancement through RKGCN comes from KGCN \cite{wang2019knowledge}. Such components enable the model to capture the knowledge graph’s high-level structural information. We can use a function g to calculate the correlation score between users and relations.

\begin{align}
\label{eq:8}
\text{score}_u^r=(u \times r^t)
\end{align}

Among them, $u \in \mathbb{R}^d$ and $r \in \mathbb{R}^d$ represent the user embedding and the relation embedding, respectively. The neighbor information of the current item can be expressed as $N_u^v$, and 

\begin{align}
\label{eq:9}
 {
N_u^v=\sum_{e \in  {N{(v)}}}{\bf e} \frac{\exp{(\text{score}_u^{r_{v,e}})}}{\sum_{e \in N(v)} \exp{(\text{score}_u^{r_{v,e}}})}.
}
\end{align}
 {where ${\bf e}$ represents the entity $e$. The right term is the normalization of user-relation score $\text{score}_u^{r_{v,e}}$, and $r_{v,e}$ denotes the relation between entities $v$ and $e$. }

The next step is to combine neighbor information with candidate item representations:

\begin{align}
\label{eq:10}
i_{h_1}=W(v+N_u^v)+b
\end{align}

\subsection{Learning Algorithm}
Through user preference aggregation and knowledge graph entity enhancement, more accurate user preferences and item embedding have been calculated, and the next step is prediction. In this paper, the prediction function f is:

{
\begin{align}
\label{eq:11}
\hat{y}_{u,v}=\sigma(i_{h_n}^t o_n).
\end{align}
}

Because the RKGCN model needs to traverse all user-item pairs, to improve efficiency, negative sampling is performed on items that the user did not interact with at a ratio of 1 : 1 with the number of interactions during selection. {The loss function is illustrated as follows}:

{
\begin{align}
\label{eq:12}
\mathcal{L}=\sum_{u \in \mathcal{U}}(\sum_{v:y_{u,v}=1}\mathcal{C}(y_{u,v}-\hat{y}_{u,v}) - \sum_{i=1}^{T_u}E_{v_i}\mathcal{C}(y_{u,v}-\hat{y}_{u,v})) + \lambda\lVert\mathcal{C}\rVert_2^2,
\end{align}
where $T_u$ and $C$ denote the number of negative samples and the cross-entropy function, respectively. $\lVert\mathcal{C}\rVert_2^2$ indicates the L2-regularizer of $\mathcal{C}$.
}

\begin{algorithm}[ht]
\caption{Procedures of the proposed RKGCN model}
\label{alg:algorithm}
\begin{algorithmic}[1]
{{
\LeftComment {{\it Graph convolutional network definition}}
\begin{itemize}
    \item{\bf Step 1:} Create the user-item interaction matrix $Y$ from the implicit interaction information by Eq. \ref{eq:1}.
    \item{\bf Step 2:} Predict the probability of a user's interaction with an item by Eq. \ref{eq:2}.
\end{itemize}
}
    
\LeftComment {{\it User preference aggregation}}
\begin{itemize}
    \item{\bf Step 1:} Obtain the triples contained in the user preference set by Eq. \ref{eq:3}.
     {
    \item{\bf Step 2:} Get the similarity between the items in the user’s historical click record and the candidate item by Eq. \ref{eq:4}.
    }
    \item{\bf Step 3:} Compute the similarity between the $h$ entities in the preference set and the candidate item in the space $R$ by Eq. \ref{eq:5}.
    \item {\bf Step 4:} Get the user representation in the first preference set by $t$ by Eq. \ref{eq:6}.
    \LeftComment {{\it User embedding}}
    \item {\bf Step 5:} Calculate the user embedding in the $H$ preference set repeatedly and aggregate with the candidate item by Eq. \ref{eq:7}.
\end{itemize}

\LeftComment {{\it Entity enhancement}}
\begin{itemize}
    \item{\bf Step 1:} Calculate the similarity between the current user and relations by Eq. \ref{eq:8}.
    \item{\bf Step 2:} Calculate the neighborhood information by Eq. \ref{eq:9}.
    \LeftComment {{\it Item embedding}}
    \item{\bf Step 3:} Aggregate multiorder neighborhoods by Eq. \ref{eq:10}.
\end{itemize}

\LeftComment {{\it Predicted probability}}
\begin{itemize}
    \item Use the inner product of item and user representations to calculate probabilities by Eq. \ref{eq:11}.
\end{itemize}

\LeftComment {{\it Loss function}}
\begin{itemize}
    \item Update the model with the loss function of Eq. \ref{eq:12}. 
\end{itemize}
}
\end{algorithmic}
\end{algorithm}

\section{Experiment}
\label{sec:exp}
In this section, we introduce the evaluation datasets and parameter settings in our experiments.

\subsection{Datasets}
{To demonstrate the effectiveness of the proposed RKGCN approach in the recommendation system}, we include the following 3 real-world datasets. Table~\ref{tab:datasets} presents the properties of each dataset.

\begin{itemize}

\item \textbf{MovieLens-1M}\footnote[1]{https://grouplens.org/datasets/movielens/1m/}~\cite{wang2018ripplenet} {is a benchmark dataset typically used for movie recommendation research. It contains approximately one million explicit ratings (on a scale of 1 to 5) of movies provided by the MovieLens website.} 
    
\item \textbf{Booking-crossing}\footnote[2]{http://www2.informatik.uni-freiburg.de/~cziegler/BX/} consists of 1,149,780 ratings from 278,858 users for 271,379 books.
    
\item \textbf{LFM-1b 2015}\footnote[3]{http://www2.informatik.uni-freiburg.de/~cziegler/BX/} is a music dataset that includes information about {artists, their albums, tracks, users, and their individual listening events. It comprises approximately three million explicit ratings for 15,471 items recorded by 12,134 users.}

\end{itemize}

\begin{table}[!tb]
\renewcommand{\arraystretch}{1.5}
\centering
\caption{{Some statistics for the three datasets used in the experiments}.}
  
\begin{tabular}{c|ccc}\hline
& MovieLens-1M & Booking-crossing & LFM-1b 2015 \\\hline    
Users & 6036 & 17860 & 12134 \\
Items & 2445 & 14967 & 15471 \\
Interaction & 753772 & 139746 & 2979267 \\\hline
Knowledge Graph Triples & 1240995 & 151500 & 464567 \\\hline
\end{tabular}
\label{tab:datasets}
\end{table}

\subsection{Baseline Models}
{We compared the RKGCN model with some baseline models to prove the performance.}

\begin{itemize}
\item \textbf{CKE}~\cite{zhang2016collaborative} combines collaborative filtering with structural, textual and visual knowledge to learn an implicit representation of users and items for recommendations.
    
\item \textbf{MCRec}~\cite{hu2018leveraging} is a recommendation model that uses metapaths, which are sequences of relationships between different entities, to represent context. MCRec uses a {common-attention mechanism to train the representations of the context}, users, and items based on their relationships with one another.
    
\item \textbf{RippleNet}~\cite{wang2018ripplenet} uses a memory network-like approach to represent users by their related items. It propagates the user's representation through all relevant entities in a knowledge graph to make recommendations. {RippleNet personalizes recommendations for users by considering the relationships between users and other entities in the knowledge graph.}
    
\item \textbf{DKN}~\cite{wang2018dkn} uses a knowledge graph to incorporate external knowledge into the recommendation process. In addition, {DKN learns the representations of the context, users, and items based on their relationships using a multichannel attention mechanism}, and uses these representations to make recommendations. DKN is “knowledge-aware" {because it uses the relations among the entities of a knowledge graph to generate recommendations}, rather than relying solely on user-item interactions.
    
\item \textbf{KGCN}~\cite{wang2014knowledge} uses a graph convolutional network (GCN) to incorporate high-order neighbourhood information from a knowledge graph into the recommendation process. {In addition, KGCN attends to the different relationships in a knowledge graph by using the representations of users to determine which neighborhoods are most likely to be of interest to them. These weights are then used to influence the recommendation process, which implies that KGCN considers the relationships between the user and other entities in the knowledge graph to produce recommendations.

 {
\item \textbf{Cn-RippleNet}~\cite{luo2021recommended} considers the influence of the entity weight of RippleNet in graph network.

\item \textbf{GMCF}~\cite{su2021neural} is a neural network graph matching-based collaborative filtering model that can effectively capture both attribute interactions by modelling and aggregating them in the graph matching structure.

\item \textbf{KANR}~\cite{zhang2021knowledge} is an attention aggregation network augmented with knowledge graphs for recommendation. It has three main components. First, the attention aggregation network collects users' interaction histories and captures the preferences of users for each item. Then, the knowledge graph-based embedding model is used to integrate knowledge. The semantic information of edges and nodes is embedded to a latent vector space. Finally, the features of the two vectors are fused to information interaction unit for recommendation. KANR-K and KANR-A are trained only
using recommendation model and average aggregation algorithm, respectively.

\item \textbf{CKAN}~\cite{wang2020ckan} is a collaborative knowledge-aware attention model that is used to distinguish the contribution of different neighborhood entities in processing knowledge graphs and cooperation information.

\item \textbf{ATKGRM}~\cite{zhang2022knowledge} is an adversarial training-based knowledge graph recommendation network. ATKGRM can adjust the aggregation weights of the knowledge graph based on adversarial training and thus learn the features of users and items more rationally.
}
}

\end{itemize}

\subsection{Experiment Setup}
{The hop number of the RKGCN is set to $H = 2$, for entities in the knowledge graph, too many hops may cause noise. The setting of hyperparameters is introduced in Table~\ref{para}. Among them, $d$ represents the dimensions of users and relations in the model. $H$ denotes the number of hops of user preference dissemination in the knowledge graph. $N_p$ and $N_e$ represent the number of elements in the user preference set and the entity neighbour set respectively. 
$iter$ is the number of aggregations in the GCN. Finally, $\lambda$ and $l_r$ represent the parameters and learning rate of $l_2$ regularization, respectively. In addition, for the division of the dataset, we divide the dataset into training, validation and test sets in the ratio of 6:2:2}. At the same time, when reporting the results, each model displays the average of the best results in 5 runs.

{
\subsection{Evaluation Measures}
We use the area under a curve with Riemann sums (AUC) and accuracy (ACC) as  indicators to evaluate the performance of the proposed recommendation system, respectively.

\begin{itemize}
\item \textbf{AUC:} AUC is often defined as the area surrounded by the axes under the ROC curve. Since the ROC curve is over the function y=x in general, the AUC has values ranging between 0.5 and 1. Moreover, the larger the AUC the better the classification performance.

\item \textbf{ACC:} ACC indicates the ratio of samples in which the predicted values are consistent with the true values. This indicator indicates the ratio of true positives and false negatives accurately identified by the classifier. The calculation is shown as follows:

\begin{align}
\label{eq:auc}
\text{ACC} = \frac{TP + TN}{TP + FP + FN + TN},
\end{align}
where TP (true positive) and FP (false positive) refer to samples with predicted and true values that are both positive, and samples with predicted values that are positive and true values that are negative, respectively. FN (false negative) and TN (true negative) refer to samples with predicted values that are negative and true values that are positive, and samples with predicted and true values that are both negative, respectively.
\end{itemize}

}

\begin{table}[!tb]
\renewcommand{\arraystretch}{2}
\centering
\caption{Parameter setting of RKGCN}
\begin{tabular}{l|l}
\hline
Parameter & Search ranges\\\hline
Moive-Lens 1M & {$d = 8, N_p = 64, N_e = 8, \lambda = 1\mathrm{e}-7, lr = 1\mathrm{e}-2$}\\
Book Crossing & {$d = 4, N_p = 16, N_e = 8, \lambda = 1\mathrm{e}-4, lr = 1\mathrm{e}-2$}\\
LFM-1b 2015 & {$d = 32, N_p = 32, N_e = 4, \lambda = 1\mathrm{e}-8, lr = 1\mathrm{e}-2$}\\\hline

\end{tabular}
\label{para}
\end{table}

\subsection{Results}
The results of the comparison test with other models are shown in Table~\ref{tab:result}.  RKGCN performed well in both AUC and ACC evaluation indicators.

\begin{table}[]
\renewcommand{\arraystretch}{2}
    \centering
    \caption{AUC and ACC results in CTR prediction.  {`-' means no result in their original papers.}}
    \begin{tabular}{lcccccc}
    \toprule
    \textbf{Model}  & \multicolumn{2}{c}{\textbf{Movie-Lens 1M}} & \multicolumn{2}{c}{\textbf{Book Crossing}} & \multicolumn{2}{c}{\textbf{LFM-1b 2015}} \\ 
     & AUC & ACC & AUC & ACC & AUC & ACC \\
     \midrule
     \textbf{CKE} & 0.800 & 0.739 & 0.671 & 0.635 & 0.903 & 0.856 \\
     \textbf{MCRec} & 0.897 & 0.826 & 0.648 & 0.619 & 0.892 & 0.843 \\
     \textbf{DKN} & 0.656 & 0.589 & 0.621 & 0.599 & 0.803 & 0.781 \\
     \textbf{KGCN} & 0.909 & 0.834 & 0.655 & 0.623 & 0.804 & 0.729 \\
     \textbf{RippleNet} & 0.920 & 0.843 & 0.660 & 0.626 & 0.913 & 0.859\\
      {\textbf{Cn-RippleNet}} & 0.920 & 0.846 & 0.745 & 0.681 & - & - \\
      {\textbf{GMCF}} & 0.918 & 0.845 & \textbf{0.789} & 0.712 & 0.785 & 0.710 \\
      {\textbf{KANR-K}} & 0.903 & 0.826 & 0.718 & 0.668 & 0.782 & 0.742 \\
      {\textbf{KANR-A}} & 0.839 & 0.826 & 0.726 & 0.679 & 0.808 & 0.746 \\
      {\textbf{CKAN}} & - & - & 0.753 & 0.675 & 0.840 & 0.772 \\
      {\textbf{ATKGRM}} & - & - & 0.782 & \textbf{0.723} & 0.857 & 0.776 \\
     \midrule
     \textbf{RKGCN} & \textbf{0.926} & \textbf{0.851} & 0.669 & 0.639 & \textbf{0.919} & \textbf{0.862}\\
     \bottomrule
    \end{tabular}
    \label{tab:result}
\end{table}

\begin{itemize}
    
\item In both the movie database and the music database, RKGCN has achieved the best performance, indicating that if the knowledge graph is sufficiently dense, {the features in the knowledge graphs are sufficiently} utilized to obtain more accurate recommendations.
    
 {\item GMCF \cite{su2021neural} and ATKGRM \cite{zhang2022knowledge} have achieved better results than other baseline models on AUC and ACC metrics in the Book-Crossing dataset, respectively. The experimental results prove that} when the knowledge graph is not dense enough, the item’s semantic information and image information can better help the recommender to make personalized recommendations. 
    
\item The DKN \cite{wang2018dkn} model achieved poor results, which proved that for recommended scenes such as movies, books, and music, the knowledge graph’s semantic information is not as crucial as its structured information.
    
\item The comparison of RKGCN with KGCN \cite{wang2019knowledge} and RippleNet \cite{wang2018ripplenet} shows that it is sufficient to enhance the representation of users and items simultaneously on the premise that the knowledge graph is sufficiently dense.

\end{itemize}


\subsection{Parameter Sensitivity}
{The effects of the hyperparameters of the RKGCN are discussed in this section}. After the experiments, we found some more important parameters, which we will explain in turn. Figure 5.5 shows the effect of parameter d on the model. In the experiment, d was changed from 2 to 32 {and the performance of the RKGCN increased first and then decreased and reached the best performance at d=8}. Figure 5.5 discusses the impact of the change in dimensions on the book database. It can be seen that both ACC and AUC first increase and then decrease.  Table~\ref{tab:auc} shows the influence of variable $N_p$ in the evaluation index AUC on the experimental results in the three databases. Table~\ref{tab:acc} shows the change in the ACC of experimental results under different $N_p$.

\begin{table}[]
\renewcommand{\arraystretch}{2}
    \centering
    \caption{The AUC result of different $N_p$ value}
    \begin{tabular}{lcccc}
    \toprule
      & $N_p$ = 8 & $N_p$ = 16 & $N_p$ = 32 & $N_p$ = 64 \\
    \midrule
    \textbf{Movie-lens 1M} & 0.918 & 0.921 & 0.925 & \textbf{0.926}\\
    \textbf{Book-Crossing} & 0.665 & \textbf{0.669} & 0.659 & 0.657\\
    \textbf{LFM-1b 2015} & 0.909 & 0.916 & \textbf{0.919} & 0.918 \\ \bottomrule
    \end{tabular}
    \label{tab:auc}
\end{table}

\begin{table}[]
\renewcommand{\arraystretch}{2}
    \centering
    \caption{The ACC result of different $N_p$ value}
    \begin{tabular}{lcccc}
    \toprule
      & $N_p$ = 8 & $N_p$ = 16 & $N_p$ = 32 & $N_p$ = 64 \\
    \midrule
    \textbf{Movie-lens 1M} & 0.843	& 0.848 & 0.851 & \textbf{0.851}\\
    \textbf{Book-Crossing} & 0.612 & \textbf{0.639} & 0.622 & 0.614\\
    \textbf{LFM-1b 2015} & 0.856 & 0.857 & \textbf{0.862} & 0.861 \\ \bottomrule
    \end{tabular}
    \label{tab:acc}
\end{table}

\begin{figure}[t]
\centering
\begin{subfigure}[AUC values in the MovieLens-1M dataset.]{
\includegraphics[width=0.47\hsize]{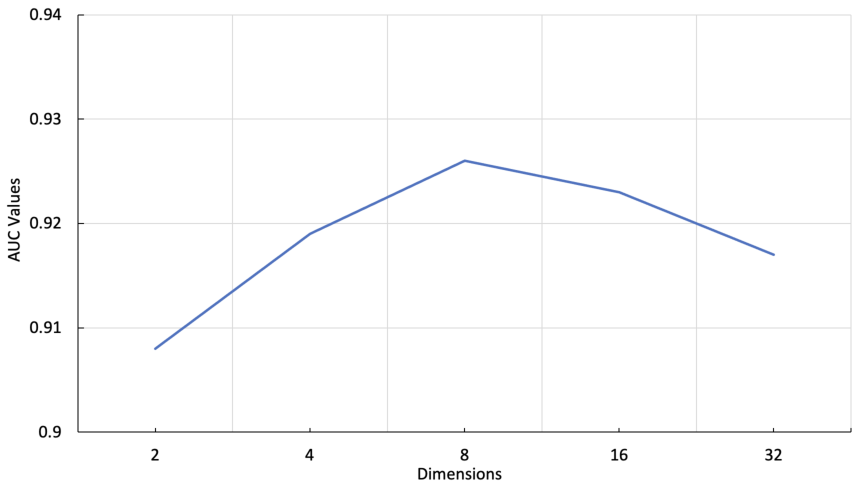}
}
\end{subfigure}
\begin{subfigure}[ACC values in the MovieLens-1M dataset.]{
\includegraphics[width=0.47\hsize]{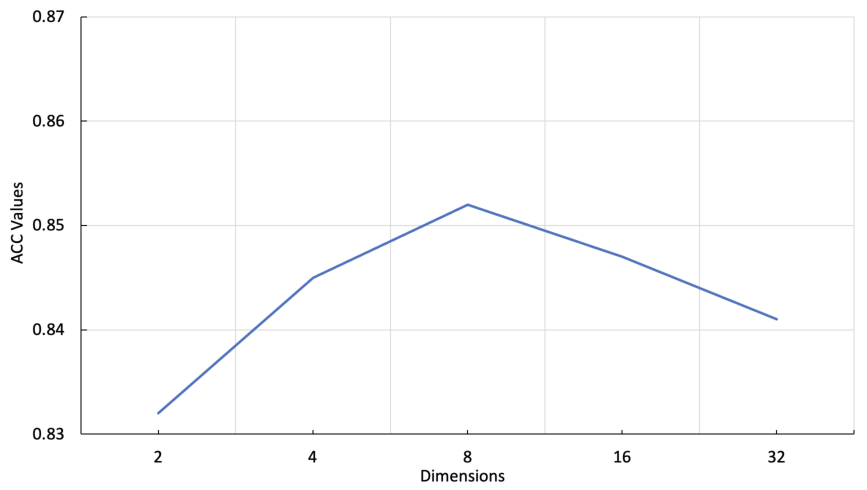}
}
\end{subfigure}
\caption{{AUC and ACC values in different dimensions in the MovieLens-1M dataset.}}
\label{fig:movielens}
\end{figure}

\begin{figure}[t]
\centering
\begin{subfigure}[AUC values in the Book-crossing dataset.]{
\includegraphics[width=0.47\hsize]{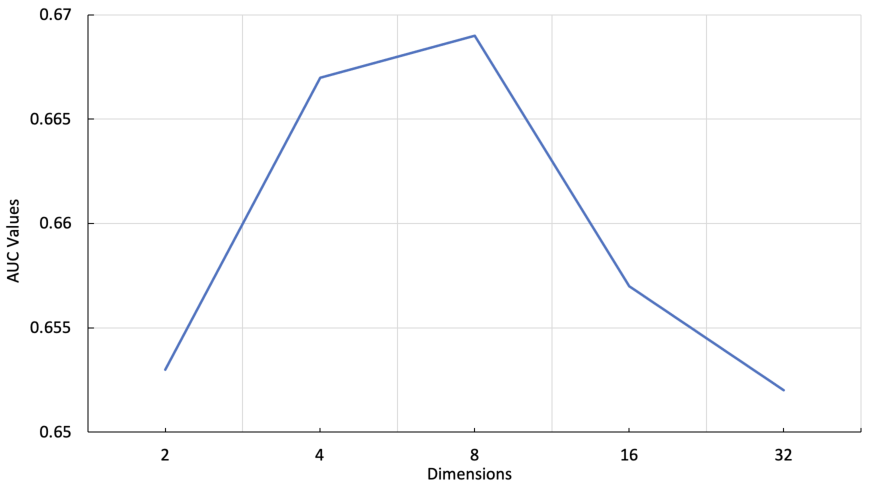}
}
\end{subfigure}
\begin{subfigure}[ACC values in the Book-crossing dataset.]{
\includegraphics[width=0.47\hsize]{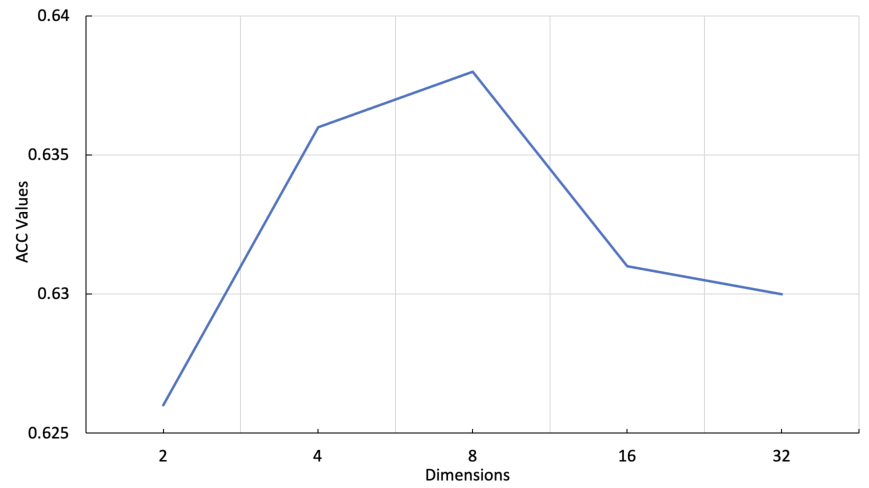}
}
\end{subfigure}
\caption{{AUC and ACC values in different dimensions in the Book-crossing dataset.}}
\label{fig:bookcrossing}
\end{figure}

\section{Conclusion}
\label{sec:conclusion}
This paper proposes an end-to-end deep learning model, i.e., RKGCN, which can effectively combine knowledge graph information to make personalized recommendations for users. RKGCN combines the knowledge graph as side information and adds the item’s information to the calculation process of user representation. We also applied RKGCN to three real scenarios, and the experimental results proved that RKGCN is better than most most of the baseline models in the two evaluation indicators.  

It is worth mentioning that RKGCN randomly samples elements in the user preference set and uses two different structures to optimize user representation and item representation. Since users and items can grow exponentially, a dataset may have millions of nodes and edges, and their feature vectors have ultrahigh dimensionality. Applying traditional GNNs for user-personalized recommendations will be challenging on large datasets. Thus, we could design a scalable GNN to handle large-scale knowledge graphs with a more effective sampling method to reduce the training graph size. Moreover, graph-based recommendation systems are susceptible to noise interference in knowledge graphs, leading to significant performance degradation. In future work, we will utilize adversarial learning to enhance the robustness of graph-based recommendation systems and design a unified system to enhance user and item features with fewer model parameters.




\bibliography{reference}

\end{document}